# Strategic nomadic-colonial switching: Stochastic noise and subsidence-recovery cycles

Jin Ming Koh[1] and Kang Hao Cheong[1,*]

[1]Engineering Cluster, Singapore Institute of Technology, 10 Dover Drive, S138683, Singapore
[*]Corresponding Author: Kanghao.Cheong@SingaporeTech.edu.sg

**Abstract**

Previously, we developed a population model incorporating the Allee effect and periodic environmental fluctuations, in which organisms alternate between nomadic and colonial behaviours. This switching strategy is regulated by biological clocks and the abundance of environmental resources, and can lead to population persistence despite both behaviours being individually losing. In the present study, we consider stochastic noise models in place of the original periodic ones, thereby allowing a wider range of environmental fluctuations to be modelled. The theoretical framework is generalized to account for resource depletion by both nomadic and colonial sub-populations, and an ecologically realistic population size-dependent switching scheme is proposed. We demonstrate the robustness of the modified switching scheme to stochastic noise, and we also present the intriguing possibility of consecutive subsidence-recovery cycles within the resulting population dynamics. Our results have relevance in biological and physical systems.

## 1 Introduction

The temporal or spatial diversification of available resources is an instinctive survival strategy amidst inexorably disadvantageous conditions. Indeed, it is known that diversity in behavioural traits and phenotypic expression is conducive for the proliferation of ecological populations[1,2]. Stochastic switching between phenotypes, intrinsically emergent from genetic pathways, can also enhance resilience towards exogenous environmental fluctuations[3,4]. Alternate migrations is also observed to enable population persistence, even when constrained to sink habitats only[5,6].

Such mechanisms are reminiscent of Parrondo's paradox. The flashing Brownian ratchet is the process that motivated Parrondo's paradox, in which two losing games are combined to produce a winning outcome[7–9]. In these paradoxes, stochastic perturbations from one game enable the exploitation of asymmetry in the other, to result in sustained capital growth[10,11]. To date, the Parrondo framework has been exceedingly valuable in understanding a wide range of physical phenomena and processes, including bulk drifts in granular and diffusive flow[12,13], chaos control via mixing of two or more chaotic processes[14–16], and entropy in information thermodynamics[17–19]. Quantum variants of Parrondo's games, which incorporate entanglement and interference effects, promise applicability in quantum computing[20–22]. Switching problems have also been analyzed under the framework[23,24], with a class of parameter switching algorithms yielding generalizations of Parrondo's paradox[25]. In biology, the Parrondo effect has been relevant in explaining transitions between unicellular and multicellular life[26], sensor evolution[27], epistatic genetic selection[28], and tumour growth dynamics[29]. There have also been many studies exploring the fundamental mathematical properties of the paradox[30–33]. It is hence apparent that the Parrondo's paradox has broad implications across many disciplines.

Recently, the periodic alternation of ecological populations between nomadic and colonial behaviours has been analyzed in the context of the paradox[34,35]. The nonlinear mechanics implemented accounts for both the Allee effect[36,37] and the potential presence of environmental fluctuations[38,39], the latter via sinusoidal noise functions coupled to the environmental carrying capacity. It was demonstrated that alternating between nomadism and colonialism could result in population persistence, despite each strategy being losing individually. These paradoxical survival scenarios occur, as long as colonies grow sufficiently quickly when resources are abundant, and switch sufficiently fast to nomadic lifestyle before resource levels become dangerously low[34,35].

In the present study, we investigate the effects of numerous types of stochastic environmental noise, and generalize the theoretical framework to accommodate realistic environmental depletion by both the colonial and nomadic sub-populations. For instance, weather patterns are known to be chaotic[40–42], and inter-species interference is inherently uncertain, dependent upon complex multi-agent dynamics[43,44]. This motivates the use of stochastic noise in place of sinusoidal noise in our theoretical framework, so that a wider range of fluctuation conditions can be modelled. We demonstrate the robustness of the reformulated switching rules to stochastic noise, and present the previously unexplored phenomenon of consecutive population subsidence-recovery cycles, thus closing the remaining lacuna in the analysis of time-based nomadic-colonial strategic switching.

## 2 Population Model

The formalism used in this paper closely follows those given in the original model[34,35]. We consider the co-existence of two sub-populations: the nomadic organisms, and the colonial ones. Their population sizes can be modelled as

$$\frac{dn_i}{dt} = g_i(n_i) + \sum_{j \neq i} s_{ij} n_j - \sum_{j \neq i} s_{ji} n_i \qquad (1)$$

where $n_i$ is the size of sub-population $i$, $g_i$ is the function describing the growth rate of $n_i$ in isolation, and $s_{ij}$ is the rate of switching to sub-population $i$ from sub-population $j$.

### 2.1 Nomadism & Colonialism

We let $n_1$ and $n_2$ be the nomadic and colonial population sizes respectively. In the absence of behavioral switching, the nomadic growth rate can be written as

$$g_1(n_1) = -r_1 n_1 \qquad (2)$$

where $r_1$ is the nomadic growth constant. We model nomadism as a losing strategy by setting $r_1 > 0$, such that $n_1$ decays with time. This is consistent with field observations that nomadic phases seldom support population growth, as evidenced by a study on slime mold *Dictyostelium discoideum*[45], dimorphic fungi *Candida albicans*[46], and jellyfish *Aurelia aurita*[47].

On the other hand, colonial population dynamics can be modelled using a modified logistic equation which incorporates the Allee effect[36,37]. Letting $r_2$, $K$ and $A$ denote the colonial growth constant, environmental carrying capacity and the critical Allee capacity respectively, we may write the colonial growth rate in isolation as

$$g_2(n_2) = r_2 n_2 \left(\frac{n_2}{\min(A, K)} - 1\right)\left(1 - \frac{n_2}{K}\right) \qquad (3)$$

Setting $r_2 > 0$, we have a positive growth rate when $A < n_2 < K$, and a negative growth rate otherwise. The $\min(A, K)$ term ensures that when $K < A$, $g_2$ is always zero or negative, as would be realistically expected. It has been shown in our previous study[34,35] that $A > 1$ always results in extinction. In other words, both the nomadic and colonial lifestyles are losing individually.

The carrying capacity $K$ can be expected to decline at a rate dependent upon the colonial and nomadic population sizes $n_1$ and $n_2$, due to resource depletion. Importantly, we add a stochastic noise function $f(t)$, to account for environment fluctuations. The rate of change of $K$ can then be expressed as

$$\frac{dK}{dt} = \alpha - \beta_1 n_1 - \beta_2 n_2 + f(t) \qquad (4)$$

where $\alpha > 0$ is the default growth rate of $K$, and $\beta_i > 0$ is the per-organism rate of habitat destruction. Without loss of generality, we may scale all variables such that $\alpha = \beta_2 = 1$.

### 2.2 Stochastic Environmental Noise

Setting $f(t) = \sum_{i=1}^{n} \gamma_i \sin \omega_i t$ with $\gamma_i, \omega_i \in \mathbb{R}$ yields the periodic noise functions proposed in the previous study[35]. As discussed in Section 1, we use non-periodic stochastic noise models in the present study, in order to capture a wider range of environmental fluctuations.

A natural extension is to replace the sinusoidal sum with an infinite series, with amplitude coefficients being normally distributed with some variance $\sigma^2$. In other words, we let $f(t) = \sum_{i=1}^{\infty} \gamma_i \sin \omega_i t$ with $\gamma_i \sim \mathcal{N}(0, \sigma^2)$. Interpreted as a Fourier series, this is equivalent to the canonical Gaussian white noise with associated variance $\sigma^2$. Notably, statistics based upon white noise are used in phylogenetics to analyze evolutionary pathways[48,49]; and white noise is similarly relevant in the theoretical understanding of ecological dynamics[50,51]. The use of white noise in our population model is therefore biologically motivated.

It is characteristic of white noise to have a flat power spectral density—that is, the power contained in every frequency interval is constant. However, there is existing evidence that fluctuations in ecological systems is better modelled using noise with power spectral densities inversely proportional to frequency[52,53]. Such a type of stochastic noise is termed $1/f$ noise, or pink noise. The relevance of $1/f$ noise extends far beyond ecology, to human cognition[54], chaos analysis[55], computer networks[56], microelectronics and quantum information[57], physical self-organizing systems[58], and even to gravitational-wave astronomy[59]. In the present study we consider $1/f$ noise in addition to white noise, along with a third combined noise model, consisting of superimposed white and $1/f$ noise with variance renormalization of $\sigma^2/2$ each.

### 2.3 Behavioural Switching

Biological clocks control an exceedingly wide range of physiological activities within animals, including sleep patterns, feeding behaviour, hormone release, and mating interactions[60–62]. It can be expected that nomadic-colonial transitions are driven in a similar fashion. As in the original model[35], we have defined a time-based switching scheme of period $T$ in terms of $t \mod T$ and switching constant $r_s > 0$:

$$s_{12} = \begin{cases} r_s & \text{if } c_2 \leq t \mod T < c_3 \\ 0 & \text{otherwise} \end{cases}$$
$$s_{21} = \begin{cases} r_s & \text{if } 0 \leq t \mod T < c_1 \\ 0 & \text{otherwise} \end{cases} \quad (5)$$

As a way of example, the phase boundaries $0 < c_1 < c_2 < c_3$ were taken to be $c_1 = C_1/K$, $c_2 = C_2/K$, and $c_3 = T = 1$ previously[35]. Real-world ecological populations indeed respond to changes in environmental resources, as the proposed scheme suggests. The sensing of environmental capacity $K$ by these populations represents a possible scenario. On the other hand, organisms can detect their population sizes with relative ease too. For instance, it is known that ant colonies track sub-population sizes to facilitate the delegation of roles, via a pheromone mechanism[63,64]; a large variety of animals are also known to adjust feeding and mating behaviours in response to population size fluctuations[65,66]. We therefore propose $c_1 = C_1/(n_1 + n_2)$ and $c_2 = C_2/(n_1 + n_2)$ as viable phase boundaries, which can realistically be achieved in real-world ecological populations. For illustrative purposes, we use $C_1 = 0.3$ and $C_2 = 0.7$ for several examples shown in this paper.

## 3 Results

Numerical simulations implementing the population model and time-based switching scheme were performed in *Mathematica 11* with the included non-linear differential system solver, which implements a generalized Implicit Differential-Algebraic Solver package[67,68]. Runge-kutta pairs of adaptive order are invoked when the differential system is non-stiff; a backward differentiation formula scheme is invoked when stiffness is detected. The generation of white and $1/f$ noise was also implemented using native functions.

### 3.1 Stochastic Environmental Noise

We present in Figure 1 simulation results for all three stochastic noise models, at low and high amplitudes. Clearly, the new switching scheme is robust towards environmental noise, and it is important to point out that this remains so even when the noise is of comparable magnitude to the initial population sizes and $K$. Even during periods of major environmental decline, the populations respond sufficiently quickly to avoid exceeding the carrying capacity. Recovery thereafter is similarly swift.

It is notable that the extinctions observed under low-amplitude periodic fluctuations[35] no longer appears with stochastic environmental noise. A plausible explanation is that the suspected mechanisms of resonance and forced oscillations can no longer occur effectively with stochastic noise, whose power spectrum is spread thinly over a wide range of frequencies. As a result, the nomadic-colonial switching scheme appears more stable on stochastic noise.

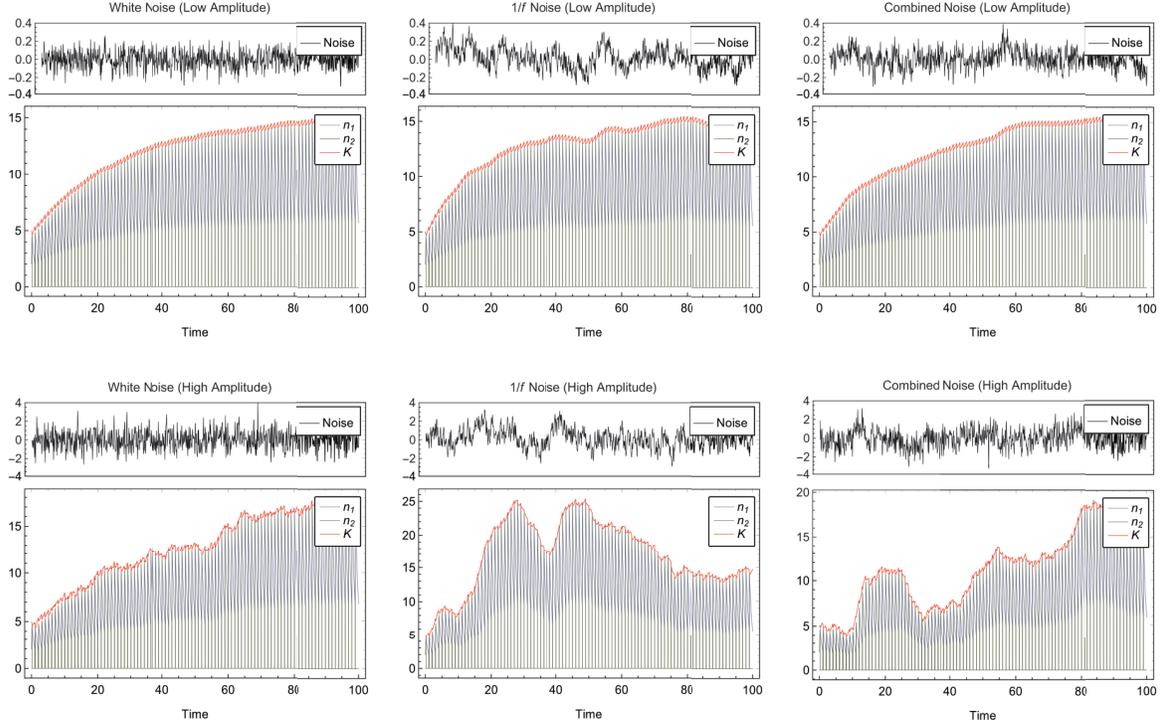

**Figure 1:** Numerical simulations of the modified time-based switching scheme incorporating population sizes, for low noise amplitude $\sigma = 0.1$ (top row) and high noise amplitude $\sigma = 1$ (bottom row). Initial conditions are $n_1 = 0$, $n_2 = 2$, $K = 5$, with $r_s = 1000$ and $\beta_1 = 0.05$. The scheme is robust towards stochastic noise.

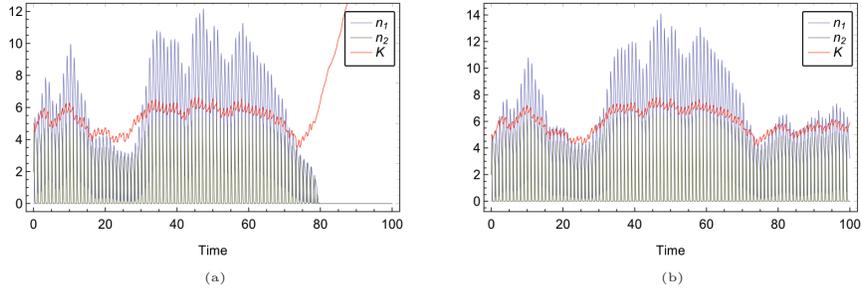

**Figure 2:** Population dynamics with switching constant $r_s = 15$ (a) and $r_s = 20$ (b). Extinction occurs when $r_s$ is insufficiently large. Initial conditions are $n_1 = 0$, $n_2 = 2$, $K = 5$, with $\beta_1 = 0.05$. Identical combined stochastic noise with $\sigma = 0.5$ is imposed on both simulation runs.

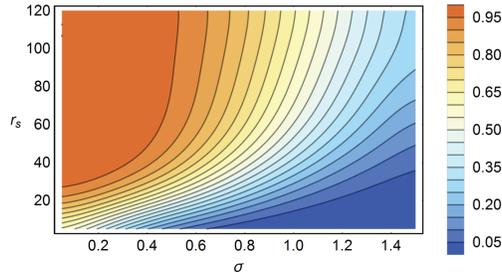

**Figure 3:** Plot of survival probability $P_s$ in $r_s$-$\sigma$ parameter space, at $t = 100$. Initial conditions are $n_1 = 0$, $n_2 = 2$, $K = 5$, with $\beta_1 = 0.05$. Numerical simulation results are averaged over 10000 repetitions.

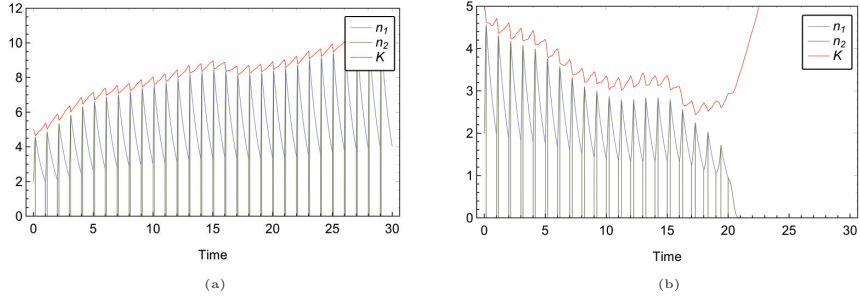

**Figure 4:** Comparison of population dynamics with $\beta_1 = 0.05$ light nomadic environmental depletion (a) and $\beta = 0.30$ heavy nomadic environmental depletion (b). Extinction occurs if the nomadic sub-population consumes environmental resources too heavily. Initial conditions are $n_1 = 0$, $n_2 = 2$, $K = 5$. Identical combined stochastic noise of $\sigma = 0.2$ is imposed on both simulation runs.

Extinction instead occurs when the switching constant $r_s$ is insufficiently large, such that the populations are not able to respond to environmental fluctuations quickly. This is demonstrated in Figure 2. Inadequate $r_s$ hinders the rate of transition between nomadic and colonial lifestyles; as a result, the population might exceed the environmental carrying capacity during periods of environmental volatility, potentially over multiple clock cycles. The consequent penalty on colonial growth rate might be so large that the population is unable to recover, hence leading to extinction.

To further demonstrate this effect, we present in Figure 3 the probability of survival $P_s$ in $r_s$-$\sigma$ parameter space. It is clearly observed that as the magnitude of environmental noise increases, steep increases in $r_s$ is necessary to ensure a reasonable chance of survival. In other words, good motility between nomadic and colonial behaviours is important to ensure survival, especially when environmental fluctuations are large, say, due to inter-species competition or volatile climate.

## 3.2 Nomadic Environmental Depletion

The coupling of both nomadic and colonial population sizes to the carrying capacity is a further development over the previous study[35]. We have demonstrated in Figure 4 the effect of nomadic environmental depletion, which is unexplored to date. We observe clearly that heavy depletion during the nomadic phase ($\beta_1 \sim \beta_2$) results in extinction.

Fundamentally, nomadic-colonial behavioural switching enables paradoxical population growth because the over-exploitation of environmental resources is avoided[34,35]. The colonial sub-population grows sufficiently fast to offset stagnation during nomadic phases, depleting the environment in the process; the species then switches to nomadism before resources dip to dangerous levels, sacrificing growth to enable the environment to regenerate. This is analogous to the agitation-ratcheting mechanism in classical Parrondo's games, in which asymmetry in one game is exploited repeatedly by activity from the other.

Any breakage in this environmental depletion-recovery cycle will disrupt sustained population growth, as is indeed seen in Figure 4. Exceedingly heavy resource consumption during the nomadic phase leaves insufficient carrying capacity for breakeven growth during the colonial phase; this debt accumulates into an eventual extinction. For survival, the nomadic lifestyle must have a depletion coefficient $\beta_1$ about an order of magnitude smaller than that of colonialism $\beta_2$.

## 3.3 Subsidence-Recovery Cycles

While the ecological relevance of the population size-dependent switching scheme has been justified in Section 2.3, it is nonetheless likely that some organisms, especially primitive ones, do not have the cognitive or communicative traits necessary for tracking population size. It is also plausible that some species choose not to utilize population size as an indicator for behavioural switching, despite having the ability to do so. In these cases, a switching mechanism based purely on biological clocks may be more relevant.

It was shown in our previous work that long-term population growth can be achieved, as long as the colonial phase decreases in duration at $t$ increases[35]. This can be made manifest via the following pure

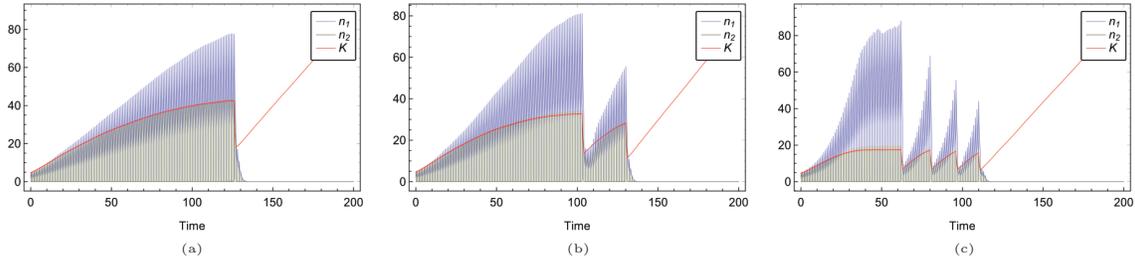

**Figure 5:** On a pure time-based switching scheme, large colonial growth rates can result in multiple subsidence-recovery cycles before eventual extinction. Comparisons between $r_2 = 30$ (a), $r_2 = 50$ (b) and $r_2 = 120$ (c) are presented. Initial conditions are $n_1 = 0$, $n_2 = 2$, $K = 5$, with $r_s = 500$ and $\beta_1 = 0.01$. Identical combined stochastic noise of $\sigma = 0.05$ is imposed on all three simulation runs.

time-based switching scheme:

$$s_{12} = \begin{cases} r_s & \text{if } c_2 \leq (1 + t/\eta)\, t \bmod T < c_3 \\ 0 & \text{otherwise} \end{cases}$$
$$s_{21} = \begin{cases} r_s & \text{if } 0 \leq (1 + t/\eta)\, t \bmod T < c_1 \\ 0 & \text{otherwise} \end{cases} \quad (6)$$

where $\eta > 0$ is a timescale constant. As a way of illustration, we use $\eta = 5$, $c_1 = c_2 = 0.15$, $c_3 = 0.45$, and $T = 1$.

Under such a switching scheme, species survival is possible with reasonable colonial growth rates ($r_2 < 20$). With larges values of $r_2$, however, over-exploitation of environmental resources occurs, resulting in species extinction. Remarkably, with even larger values of $r_2$, consecutive cycles of near-extinction and recovery is observed, with the nomadic and colonial population sizes oscillating in a sawtooth-like manner before final extinction. Figure 5 demonstrates this intriguing result. It should be noted that this phenomenon of subsidence-recovery cycles occurs both with and without environmental noise—it is an intrinsic property of the switching structure, not one induced by external noise.

The mechanism for this behaviour can be understood as follows. First, the rapid colonial growth rate results in sustained population inflation. At some point in time, a nomadic-to-colonial transition introduces such a large colonial population size that the environmental carrying capacity is greatly exceeded, thus triggering a decline in population size. As the nomadic-to-colonial transition is still ongoing, the carrying capacity remains on a declining trend. This sets up a positive feedback that forces the collapse of the species essentially within a single clock cycle, the steepness of which is indeed observed in Figure 5.

Near the final moments of collapse, the population begins the transition to nomadism, essentially allowing the species to preserve some individuals alive as colonialism plunges into extinction. This switch is aided by the temporal compression of colonial phases as $t$ increases. By the time the next switch occurs, the environmental carrying capacity has recovered, thus allowing the species to repopulate. This accounts for the recovery phase of the cycle.

Such subsidence-recovery cycles are reminiscent of the population dynamics of bacteria under the influence of antibiotics[69–71]. Upon the introduction of antibiotics, subpopulations that are not resistant quickly diminish, leaving only the resistant strains preserved alive. These preserved resistant strains can then repopulate when conditions are favourable, typically after the drug has been metabolized. The antagonistic pressure from the drug mirrors the environmental carrying capacity in our context, and the diversification of bacterial phenotypes can be considered analogous to the nomadic-colonial switching strategy, with external environmental noise modelling fluctuations in the immune response of the host. It is also known, moreover, that certain types of micro-organisms can transition into non-reproductive, incredibly resilient dormant states termed as endospores[72,73]. This allows survival amidst harsh conditions, in a similar fashion to nomadic-colonial switching. The subsidence-recovery cycles observed in our theoretical framework thus have potential applications in microbiology and ecology, particularly in the quantitative modelling of population dynamics.

# 4    Conclusion

At the spatial and temporal scales relevant to ecological dynamics, non-periodic stochastic noise can allow a wider range of environmental fluctuations to be modelled. In general, these fluctuations arise from a combination of chaotic weather patterns, inter-species competition, and complex interactions within the food web. By demonstrating the robustness of the modified time-based switching strategy to stochastic environmental noise, we extend the relevance of nomadic-colonial alternation to many other real-world biological systems. Our results also highlight the importance of good behavioural motility, such that responses to environmental fluctuations are enacted sufficiently quickly; this is indeed consistent with observations in evolutionary biology[74,75].

Throughout the present study, a realistic coupling for the environmental carrying capacity involving both the nomadic and colonial population size is considered. Species survival is found to be possible, as long as environmental depletion is not excessively heavy during the nomadic phase. Importantly, we have also demonstrated the presence of consecutive subsidence-recovery cycles under a pure time-based switching scheme. This remarkable result is closely relevant to ecological and microbiological population modelling. It is worth noting that the generalizability of our framework aids greatly in extending the model into other disciplines, for instance, in the modelling of wildfire and reforestation[76,77], human resource consumption[78], and generic environmental sustainability encompassing anthropological and natural effects. These applications all involve multiple temporal phases, each presenting different coupling to resource quantities; and they are all subject to a degree of stochasticity. The relevance of the framework is similarly extendable to physical multi-body systems[79], such as self-organizing automata and swarm robotics[80,81], which may exploit switching strategies to achieve common goals.